\documentclass[aps,prd,showpacs,groupedaddress,nofootinbib,twocolumn,amsmath,amsfonts,amssymb]{revtex4} 

\usepackage{amsmath}
\usepackage{amssymb}
\usepackage{epsfig}
\usepackage{natbib}

\usepackage{graphicx}
\usepackage{stmaryrd}
\usepackage{bm}
\usepackage{color}
\usepackage{textcomp}

\newcommand{\AddrMPP}{Max-Planck-Institut f\"ur Physik (Werner Heisenberg
Institut),
F\"ohringer Ring 6, 
80805 M\"unchen, Germany}

\begin{document}

\title{Consistent analysis of the $\nu_\mu \to \nu_e$ sterile neutrinos searches of ICARUS and OPERA}

\author{Antonio Palazzo}
\affiliation{\AddrMPP}


\begin{abstract}

The two long-baseline experiments ICARUS and OPERA have recently provided bounds on light ($\sim$ eV) sterile neutrinos exploiting the results of the $\nu_\mu \to \nu_e$ appearance searches. Both collaborations have performed the data analysis using an effective 2-flavor description. We show that such a simplified treatment neglects sizable genuine 4-flavor effects, which are induced by the interference of the new large squared-mass splitting $\Delta m^2 _{14}$ with the atmospheric one.  The inclusion in the data analysis of such effects weakens the upper bounds on the effective appearance amplitude $\sin^2 2\theta_{\mu e}$ approximately by a factor of two. In addition, we point out that, in a 4-flavor scheme, the flavor oscillations involve also the $\nu_e$ component of the CNGS beam and can suppress the theoretical expectation of the background in a substantial way. The inclusion in the data analysis of the $\nu_e$ disappearance effects leads to a further weakening of the upper bounds on $\sin^2 2\theta_{\mu e}$, which overall are relaxed by a factor of three with respect to those obtained in the effective 2-flavor description.  

\end{abstract}


\maketitle

\section{Introduction}

One of the most interesting issues in present-day neutrino physics  is provided by the hints of light ($\sim\rm{eV}$) sterile species suggested by the short-baseline (SBL) anomalies (see~\cite{Palazzo:2013me,Kopp:2013vaa,Giunti:2013aea}).  In the upcoming years new experiments will check if the anomalies are artifacts or real effects. In the mean time, sterile neutrinos can be actively investigated in other contexts. A particularly important example is that of the long-baseline (LBL) experiments.  In these setups the distance-over-energy ratio $L/E$  is much bigger than that probed in the SBL ones and no information on the new large squared-mass splitting can be recovered since the new high-frequency oscillations are completely averaged out. On the other hand, distinctive 4-flavor effects can emerge exclusively at long distances, inducing modifications of the expected rate and of its energy spectrum. Such effects depend on the other oscillation parameters involved in the frameworks  with extra sterile species, i.e. the new mixing angles and  the new CP-violating phases. Notably, as first pointed out in~\cite{Klop:2014ima}, the LBL experiments are the sole setups where the new sterile-induced CP-phases can be measured. Indeed, the running experiment T2K is already providing precious hints on one of them~\cite{Klop:2014ima}. 

At the short baselines, LSND~\cite{Aguilar:2001ty} and MiniBooNE~\cite{Aguilar-Arevalo:2013pmq} have observed an anomalous $\nu_\mu \to \nu_e$ appearance signal and its (dis-)confirmation 
is one of the main targets of the sterile neutrino searches. Notably, the two experiments ICARUS~\cite{Antonello:2012pq,Antonello:2015jxa} and OPERA~\cite{Agafonova:2013xsk} have recently provided
the first bounds on the $\nu_\mu \to \nu_e$ conversion using LBL data.  Both experiments have analyzed their results using an effective 2-flavor framework. Here we show that such a treatment neglects sizable genuine  4-flavor effects induced by the interference between the new large squared-mass splitting and the atmospheric one. Their inclusion in the data analysis provides upper bounds on the effective appearance mixing angle $\theta_{\mu e}$ that are weaker approximately by a factor of two with respect to those obtained in the 2-flavor approach. In addition, we point out that, in a 4-flavor scheme, the flavor oscillations involve also the $\nu_e$ component of the CNGS beam and can suppress the theoretical expectation of the background in a substantial way. The inclusion in the data analysis of the $\nu_e$ disappearance effects leads to a further weakening of the upper bounds on $\theta_{\mu e}$, which overall are relaxed by a factor of three with respect to those obtained in the effective 2-flavor description.

\section{Theoretical Framework}

In the presence of a fourth sterile neutrino $\nu_s$, the flavor and the mass eigenstates  are connected through a $4\times4$ mixing matrix. For LBL transitions, a convenient parameterization of the mixing matrix is
\begin{equation}
\label{eq:U}
U =   \tilde R_{34}  R_{24} \tilde R_{14} R_{23} \tilde R_{13} R_{12}\,, 
\end{equation} 
where $R_{ij}$ ($\tilde R_{ij}$) represents a real (complex) $4\times4$ rotation in the ($i,j$) plane
containing the $2\times2$ submatrix 
\begin{eqnarray}
\label{eq:R_ij_2dim}
     R^{2\times2}_{ij} =
    \begin{pmatrix}
         c_{ij} &  s_{ij}  \\
         - s_{ij}  &  c_{ij}
    \end{pmatrix}
\,\,\,\,\,\,\,   
     \tilde R^{2\times2}_{ij} =
    \begin{pmatrix}
         c_{ij} &  \tilde s_{ij}  \\
         - \tilde s_{ij}^*  &  c_{ij}
    \end{pmatrix}
\,,    
\end{eqnarray}
in the  $(i,j)$ sub-block, with
\begin{eqnarray}
 c_{ij} \equiv \cos \theta_{ij} \qquad s_{ij} \equiv \sin \theta_{ij}\qquad  \tilde s_{ij} \equiv s_{ij} e^{-i\delta_{ij}}.
\end{eqnarray}
The  parameterization in Eq.~(\ref{eq:U}) enjoys the following properties: I) For vanishing mixing
with the fourth state $(\theta_{14} = \theta_{24} = \theta_{34} =0)$ 
it reduces to the 3-flavor matrix in its usual parameterization.
II) The leftmost positioning of the matrix $\tilde R_{34}$ makes the $\nu_{\mu} \to \nu_{e}$    
conversion probability independent of $\theta_{34}$ (and the related CP-phase $\delta_{34}$). 
III)  For small values of $\theta_{13}$ and of the mixing angles involving the fourth mass eigenstate, 
one has $|U_{e3}|^2 \simeq s^2_{13}$, $|U_{e4}|^2 = s^2_{14}$ (exact), 
$|U_{\mu4}|^2  \simeq s^2_{24}$ and $|U_{\tau4}|^2 \simeq s^2_{34}$, 
with a clear physical interpretation of the new mixing angles. 

Let us now come to the flavor conversions relevant for ICARUS and OPERA.
In the 3-flavor limit, the $\nu_\mu \to \nu_e$  transition probability can be written
as the sum of three distinct terms. The first one is driven by the atmospheric
splitting, the second one by the solar splitting, and the third one by their interference.
For the baseline of $L = 732$ km and the (average) energy $E  =  17$ GeV probed 
with the CNGS beam, we have $\Delta \equiv \Delta m^2_{13}L/4E \sim 0.13$ for the
atmospheric oscillating phase, while the solar one is thirty times smaller. Therefore,
only the atmospheric term is relevant  and we have
\begin{equation}
P^{3\nu}_{\mu e}  \simeq\,  4 s_{23}^2 s^2_{13}  \sin^2{\Delta}\,,
\end{equation}
which is $O(\epsilon^4)$ in the small parameters $s_{13} \simeq 0.15$ and $\Delta \simeq 0.13$,
which can be both assumed of order $\epsilon$.

In the 4-flavor case, neglecting the solar 
squared-mass splitting, the transition probability can be expressed as
the sum of three terms~\cite{Klop:2014ima}, 
\begin{eqnarray}
\label{eq:Pme_4nu_6_terms}
P^{4\nu}_{\mu e}  &\simeq&  P^{\rm{ATM}} +  P^{\rm {INT}} + P^{\rm {STR}}\,,
\end{eqnarray}
driven respectively by the atmospheric splitting, the sterile one
and by their interference. After averaging over the fast oscillations induced
by the large frequency $\Delta m^2_{14}$, we find in vacuum~\cite{Klop:2014ima} 
\begin{eqnarray} 
\label{eq:Pme_4nu_vac}
P^{4\nu}_{\mu e}  & = & c_{14}^2 c_{24}^2 P_{\mu e}^{3\nu} \\
\nonumber
&+& 4 c_{14}^2 c_{24} s_{14} s_{24} s_{13} s_{23} \sin\Delta \sin (\Delta + \delta') \\
\nonumber
&+&  2 c_{14} ^2 s_{14}^2 s^2_{24} \,,
\end{eqnarray} 
where  $\delta' \equiv \delta_{13} - \delta_{14}$.  The first term in Eq.~(\ref{eq:Pme_4nu_vac}) coincides
with the 3-flavor probability apart from the multiplying  factor $c_{14}^2 c_{24}^2 $.
The second term encodes the interference effects and can assume both positive and negative values.
The third term can be interpreted as the averaged transition probability in an
effective 2-flavor description. 

The interference term depends
on the neutrino mass hierarchy (NMH), i.e. from the sign of $\Delta$, which is positive
for normal hierarchy (NH) and negative for inverted hierarchy (IH). While for a 
fixed value of the CP-phase $\delta' \ne (0, \pi$) the interference term 
depends on the NMH, it is invariant under the simultaneous transformations  
\begin{eqnarray} 
\label{eq:phase_symm}
\Delta   \to   -\Delta, \qquad \delta' \to -\delta'. 
\end{eqnarray} 
This implies that there is a complete degeneracy among the NMH 
and the sign of the CP-phase $\delta'$.  For small values of $\Delta$,
like those involved in the CNGS beam, the following approximate proportionality relation holds
\begin{eqnarray} 
\label{eq:Pint_approx}
P^{\mathrm{INT}}  \propto \sin 2\Delta \sin \delta'\,,
\end{eqnarray} 
which implies that the amplitude of the interference term is maximal
for $\delta' \simeq \pm \pi/2$. 

Inspection of Eq.~(\ref{eq:Pme_4nu_vac}) shows that 
the interference and the sterile terms are proportional, 
respectively, to the first and the second power of the quantity  
\begin{eqnarray}
\label{eq:Appearance_angle}
\sin 2\theta_{\mu e} \equiv 2 |U_{e4}|  |U_{\mu 4}|  = 2 c_{14} s_{14} s_{24} \,,
\end{eqnarray}
which defines the effective appearance mixing angle probed in the SBL  $\nu_\mu \to \nu_e$ experiments.
Furthermore, the atmospheric and interference terms are respectively proportional to the
factor $\mathcal{F} \equiv c_{14}^2 c_{24}^2$ and to its squared root. 
Therefore, Eq.~(\ref{eq:Pme_4nu_vac}) can be recast in the form 
\begin{eqnarray}
\label{eq:Pme_app_angle}
P^{4\nu}_{\mu e}  &=&  \mathcal{F}  P_{\mu e}^{3\nu} \\
\nonumber
&+& 2  \sqrt{\mathcal{F}} \sin 2\theta_{\mu e} s_{13} s_{23} \sin\Delta \sin (\Delta + \delta')\\ 
\nonumber
&+& \frac{1}{2}  \sin^2 2\theta_{\mu e}\,.
\end{eqnarray}
In the particular case $|U_{e4}|  = |U_{\mu4}|$,  the suppression factor $\mathcal{F}$
is  a function of the sole effective appearance angle%
\footnote{It can be observed that, for a fixed value of $\theta_{\mu e}$, the factor 
$\mathcal{F}$ is always smaller than that obtained  in the case $|U_{e4}|  = |U_{\mu4}| $. 
In fact, the inequality  $(|U_{e4}| - |U_{\mu 4}|)^2 \ge 0$ implies that
$ \mathcal{F} \equiv 1- (|U_{e4}|^2 + |U_{\mu4}|^2)  \le 1- 2 |U_{e4}| |U_{\mu 4}|   \equiv  1 - \sin 2\theta_{\mu e} $.}
\begin{eqnarray}
\label{eq:Suppression_fact}
\mathcal{F} \equiv c_{14}^2 c_{24}^2  = 1 -  |U_{e4}|^2 - |U_{\mu 4}|^2  =  1 - \sin 2\theta_{\mu e} \,, 
\end{eqnarray}
and the transition probability is sensitive to the  4-flavor effects only through such an effective mixing angle.

It is now crucial to observe that the official analyses performed by ICARUS and OPERA
make use of an effective 2-flavor description, which by definition neglects the interference term
in the transition probability.
In order to determine the level of (in-)accuracy of the 2-flavor approximation, we must 
evaluate the relative size of the three terms in the conversion probability. 

\begin{figure}[t!]
\vspace*{-2.85cm}
\hspace*{-0.45cm}
\includegraphics[width=12.0 cm]{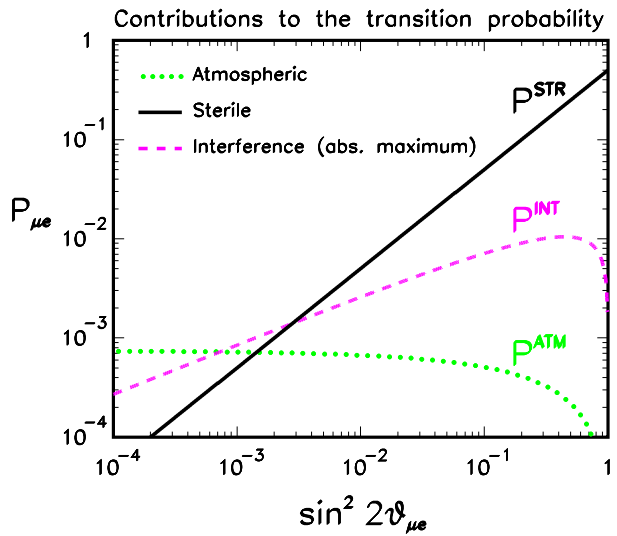}
\vspace*{-2.5cm}
\caption{Behavior of the three terms of the $\nu_\mu \to \nu_e$ transition probability for the CNGS parameters
($L = 732$~km, $E = 17$~GeV)  as  a function of $\sin^2 2\theta_{\mu e}$ in the case $|U_{e4}|  = |U_{\mu4}|$. 
\label{eq:pme_six_terms}}
\end{figure}  

Figure~1 displays their behavior as a function of $\sin^2 2\theta_{\mu e}$. For the interference term, which can have both positive and negative sign, we have plotted its maximal absolute value. In Fig.~1 we have assumed $|U_{e4}| = |U_{\mu4}|$, in which case $\theta_{\mu e}$ is the sole (4-flavor) mixing angle entering the transition probability, as explained above. For relatively small values of the two new mixing angles ($|U_{e4}|^2, |U_{\mu 4}|^2 \lesssim 0. 2$), this assumption is almost irrelevant and for whatever choice of $|U_{e4}|^2 \ne |U_{\mu4}|^2$ the plot would be almost identical. For very large values of $|U_{e4}|^2$ (or $|U_{\mu 4}|^2$) sizable deviations would appear in the atmospheric and interference terms if  $|U_{e4}| \gg |U_{\mu4}|$ (or 
 $|U_{e4}| \ll |U_{\mu4}|$). It is important to stress that, in any case, their amplitude would be always smaller than that obtained for $|U_{e4}| = |U_{\mu4}|$ displayed in Fig.~1 (see footnote 1 and discussion of Fig.~5).

As expected from Eq.~(\ref{eq:Pme_app_angle}), for small values of $\theta_{\mu e}$, the sterile term (solid line) and the interference one (dashed line) display a power-law behavior. For very large values of $\theta_{\mu e}$, the interference term deviates from the power-law behavior because  of the effect of the suppressing factor $\sqrt\mathcal{F}$, which becomes appreciably smaller than one. In the atmospheric term (dotted line), $\mathcal{F}$ is the sole source of the dependence on $\theta_{\mu e}$. This terms assumes the maximum value in the 3-flavor limit  ($\theta_{\mu e} = 0$) and decreases with increasing $\theta_{\mu e}$. In the region $\sin^2 2\theta_{\mu e} \gg 0.1$, the factor $\mathcal{F} $ becomes very small and drastically suppresses both the atmospheric and the interference terms.

For values of $s_{14}$ and  $s_{24}$ similar to that of $s_{13}$ ($\simeq 0.15$), which are favored by the SBL global fits~\cite{Kopp:2013vaa,Giunti:2013aea}, we can assume that the three mixing angles and the atmospheric oscillating phase $\Delta \simeq 0.13$ have all the same order of magnitude $\epsilon$. In this regime, corresponding roughly to $\sin^2 2\theta_{\mu e} \simeq  few \times 10^{-3}$,  all the three terms  have the same order $\epsilon^4$, and the transition probability is below the current sensitivity of the two experiments.

As can be deduced from Fig.~1, for the values of the transition probability currently probed by ICARUS and OPERA ($P_{\mu e}\sim few \times 10^{-3}$), roughly corresponding to $\sin^2 2\theta_{\mu e} \sim 10^{-2}$,  the absolute size of the interference term is comparable to (approximately one half of) the sterile term. This means that, for those values of the CP-phase $\delta'$ that render the interference term maximal and negative, the overall signal decreases by a factor of two with respect to the effective 2-flavor description. 

Figure~2  further illustrates the role of the 4-flavor effects, representing $P_{\mu e}$ as a function of the neutrino energy. The (common) value chosen for the two mixing angles ($s_{14}^2 = s_{24}^2 = 0.05$) corresponds to an effective appearance mixing angle $\sin^2 2\theta_{\mu e} \simeq 10^{-2}$,  close to the sensitivity of  the two experiments.  The thin dotted curve represents the atmospheric contribution, while the thin dotted-dashed (horizontal) line is the (energy independent) effective 2-flavor probability. The sum of these two terms, represented by the thick dashed curve, is the probability implemented in the official analyses. The two solid curves correspond to the 4-flavor transition probability calculated for the two values of the CP-phase $\delta'  = \pm \pi/2$ in the case of normal hierarchy. We see again that, in the energy region of interest, located around $[10-30]$\,GeV, the transition probability is quite different from the one used by the two collaborations. In addition, we can observe that the interference term appreciably modifies  the energy dependence of the probability. 

The discussion made above makes it clear that the inclusion of the interference effects in the analysis is expected to introduce substantial modifications of the bounds obtained in their absence. In particular, the upper limits on $\theta_{\mu e}$ should become weaker since the interference term, when negative, decreases the predicted signal. This qualitative expectation will be quantified by the numerical analysis presented in the next section.

An important remark is in order before presenting the results of the analysis. Both ICARUS and OPERA operate in a background-dominated regime and the $\nu_e$ contamination of the CNGS beam is the main (almost the sole)  source of background to the $\nu_\mu \to \nu_e$ signal. In the 3-flavor limit the $\nu_e \to \nu_e$ survival probability $P_{ee}$ is equal to one apart from negligible $O(\epsilon ^4)$ corrections. In this case, the $\nu_e$ beam component is unaffected by the oscillations and the background can be considered a fixed quantity. Differently, in a 4-flavor scheme, the survival probability $P_{ee}$ can be appreciably different from unity and one cannot assume that the $\nu_e$ background is a fixed quantity as done in the official
analyses of the two collaborations.  In the region of high values of $\Delta m^2_{14}$, where the oscillations are averaged, we have 

\begin{eqnarray}
P_{ee} \simeq 1 - 2|U_{e4}|^2 (1- |U_{e4}|^2)\,.
\label{eq:Pee}
\end{eqnarray}

It is evident that large values of $|U_{e4}|$ are expected to substantially 
suppress the background and thus decrease the sensitivity to a potential signal coming 
from the $\nu_\mu \to \nu_e$ transitions. The analysis presented in the next section will 
take this aspect into proper account.  
 
\begin{figure}[t!]
\vspace*{-2.80cm}
\hspace*{-0.53cm}
\includegraphics[width=12.0 cm]{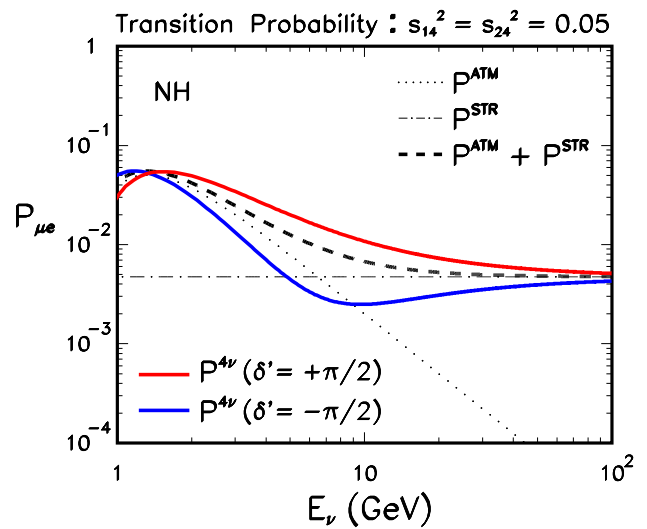}
\vspace*{-2.7cm}
\caption{Transition probability as a function of the neutrino energy. 
The dotted curve represents the atmospheric term, while the dotted-dashed (horizontal) line is 
the sterile one. The sum of these two contributions, represented by the dashed curve, is the probability implemented in the official analyses. The two solid lines correspond to the (averaged) 4$\nu$ probability in the NH case for the two values  $\delta' = \pm \pi/2$.
\label{eq:pme_six_terms}}
\end{figure}  

\section{Numerical Analysis}

In our analysis we use the results of the $\nu_\mu \to \nu_e$ appearance searches provided in~\cite{Antonello:2012pq,Antonello:2015jxa} for ICARUS and in~\cite{Agafonova:2013xsk} 
for OPERA. In order to calculate the theoretical expectation for the total number of events,
we convolve the product  of the $\nu_\mu$ flux, the cross-section, and the $\nu_\mu \to \nu_e$  
transition probability%
\footnote{We have calculated the transition probability numerically including the matter 
effects, albeit these have a negligible role.}
with the energy resolution function and the detection efficiency.
 A similar computation is performed for the $\nu_e$ beam
component, incorporating the $\nu_e \to \nu_e$ survival probability.
 We have checked that our predictions for the expected $\nu_e$ rate
are in good agreement with those published. 
For definiteness, we show the results obtained for the OPERA experiment, 
the case of ICARUS being completely analogous. To make the discussion more clear, 
in our analysis, we first consider the particular case $P_{ee} = 1$, i.e. we neglect the
oscillations of the $\nu_e$ beam component (Figs.~3 and 4). Then, we extend our analysis to
the general case $P_{ee} <1$ (Fig.~5). This way of presenting the results 
 will allow us to separate the effects on the $\nu_\mu \to \nu_e$ appearance from those,
 conceptually different, related to the $\nu_e$ disappearance.

\begin{figure}[t!]
\vspace*{-2.85cm}
\hspace*{-0.45cm}
\includegraphics[width=11.8 cm]{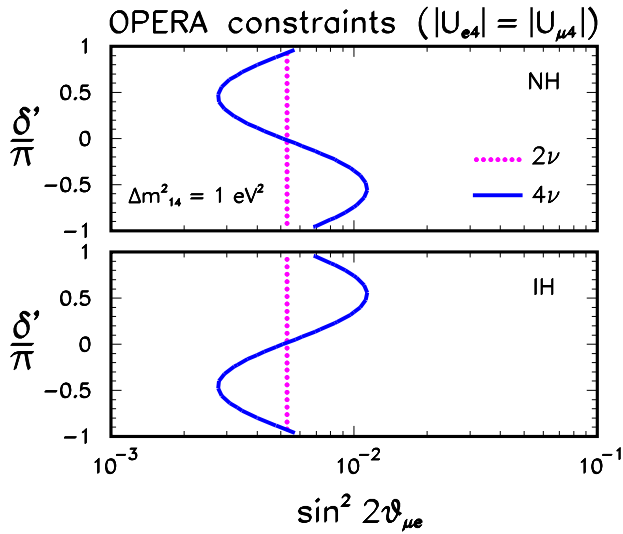}
\vspace*{-2.5cm}
\caption{Upper bounds  (90\% C.L.) obtained for a fixed (large) value of $\Delta m^2_{14}$
in the two cases of NH and IH. The effect of the oscillations on the $\nu_e$ component
is neglected setting $P_{ee} = 1$.
\label{fig:opera_fixed_dm2}}
\vspace*{0.0cm}
\end{figure}  

Figure~\ref{fig:opera_fixed_dm2} shows the 90\% C.L. upper bounds that we obtain  on the appearance mixing angle as a function of the CP-phase $\delta'$ for the OPERA experiment  in the two cases of NH (upper panel) and IH (lower panel)
for $\Delta m^2_{14} = 1$\,eV$^2$. In both panels, the dashed vertical line represents the upper bound that we obtain in the 2-flavor approximation  ($\sin^2 2\theta_{\mu e} \lesssim 5.2 \times 10^{-3}$), which is in good agreement with the  limit quoted by OPERA. The solid contour represents the upper bounds obtained in the 4-flavor scheme, assuming
 $|U_{e4}| = |U_{\mu 4}|$ and  $P_{ee} =1$. As expected, a dependence on the CP-phase $\delta'$ appears that is different in the two cases of NH and IH.  The 4-flavor upper limits are substantially stronger (weaker) than those obtained in the 2-flavor case when the interference term assumes positive (negative) values. The maximal excursion from the 2-flavor result, basically identical for NH and IH, is obtained for $\delta' \simeq \pm \pi/2$, as expected from the discussion made in Sec.~II.

\begin{figure}[t!]
\vspace*{-2.85cm}
\hspace*{-0.45cm}
\includegraphics[width=11.8 cm]{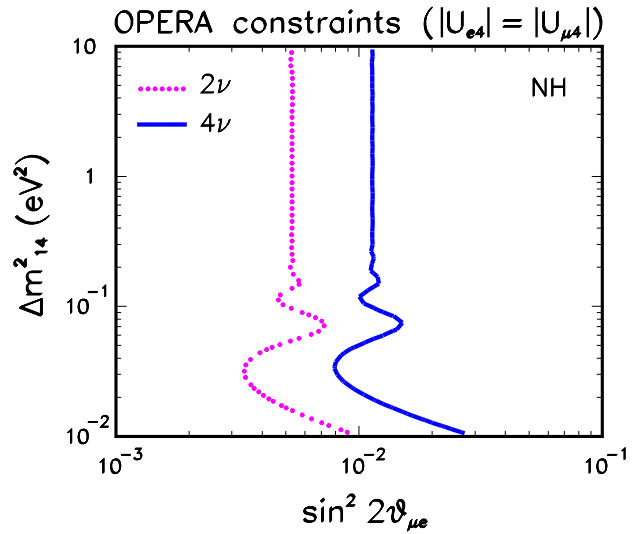}
\vspace*{-2.5cm}
\caption{Upper bounds (90\% C.L.) obtained in the case of NH. 
The CP-phase $\delta'$ is marginalized away. 
The effect of the oscillations on the $\nu_e$ component is neglected setting $P_{ee} = 1$.
\label{fig:opera_theta_mue_dm2}}
\end{figure}  

Figure~\ref{fig:opera_theta_mue_dm2} shows the upper bounds in the usual plane [$\sin^2 2\theta_{\mu e},  \Delta m^2_{14}]$. The constraints displayed for the 4-flavor case, valid for $|U_{e4}| = |U_{\mu 4}|$ and $P_{ee} =1$, correspond
to the case of NH and are obtained by marginalizing away the CP-phase $\delta'$. For IH the results (not shown) are basically identical. It is evident that, independently of the value of $\Delta m^2_{14}$, the upper limits obtained by the full 4-flavor analysis are approximately a factor of two weaker than those obtained in the 2-flavor case.  More precisely,
in the high-$\Delta m^2_{14}$ region, we obtain $\sin^2 2\theta_{\mu e}\lesssim 1.2 \times 10^{-3}$.

The upper bounds presented in  Figs.~3 and 4 have been obtained for the particular 
choice $ |U_{e4}| = |U_{\mu 4}|$. Hence, the question arises on which kind of deviations
emerges when this condition is relaxed. It can be easily seen that for any choice of 
$|U_{e4}| \ne |U_{\mu 4}|$ the bounds are stronger than those derived in the particular case 
$|U_{e4}| = |U_{\mu 4}|$. In fact,  in this last case, the amplitude of the interference term is maximized
(see footnote 1) and the deviations from the 2-flavor description are maximal. Hence, the case presented provides the weakest bounds on $\theta_{\mu e}$. 
Figure~5 visualizes this behavior, showing the constraints obtained in the general case,
i.e. when the two amplitudes  $|U_{e4}|^2$ and  $|U_{\mu 4}|^2$ are both allowed to vary 
(respecting the unitarity constraint  $|U_{e4}|^2 + |U_{\mu 4}|^2 \le1$).  
The new splitting is fixed at $\Delta m^2_{14} = 1$\,eV$^2$, while the 
CP-phase $\delta'$ is marginalized away. In the 2-flavor approximation (dotted line) 
we re-obtain the bound $\sin^2 2\theta_{\mu e} \lesssim 5.2 \times 10^{-3}$,
which in the log-log plot of Fig.~5 is represented by a diagonal line with a negative slope
of 45\textdegree\, (we recall that $\sin^2 2\theta_{\mu e} \equiv 4 |U_{e4}|^2|U_{\mu 4}|^2$).
The solid curve represents the bounds obtained when the interference term is ``switched on''
in the conversion probability. It can be clearly seen that the weakest bound is
obtained for $ |U_{e4}| = |U_{\mu 4}|$ as anticipated, in which case we re-obtain 
$\sin^2 2\theta_{\mu e} \lesssim 1.2 \times 10^{-2}$. For different choices of the mixing 
amplitudes the upper bounds exhibit  deviations that are generally small, becoming 
appreciable only if one of the two amplitudes  is very large.

\begin{figure}[t!]
\vspace*{-1.6cm}
\hspace*{-0.55cm}
\includegraphics[width=12.3 cm]{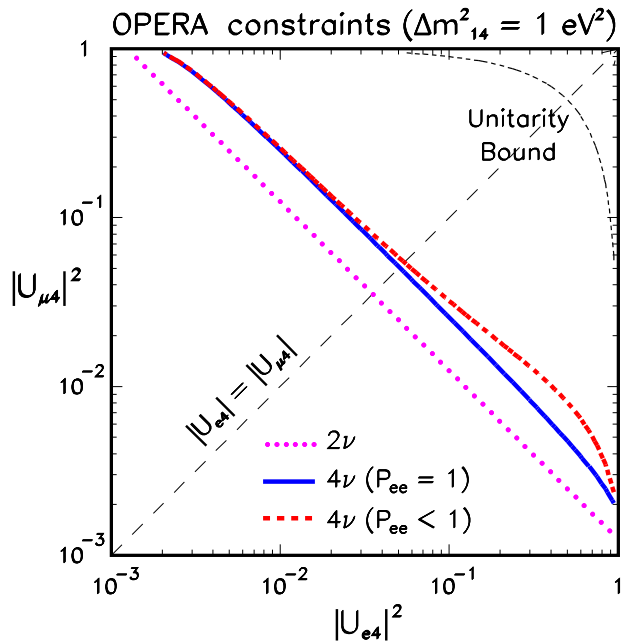}
\vspace*{-2.7cm}
\caption{Upper bounds (90\% C.L.) obtained from the OPERA experiment for the case of normal hierarchy.
The CP-violating phase $\delta'$ is marginalized away. See the text for details. 
\label{fig:opera_ue4_um4}}
\end{figure}  

Let us now come to the impact of the oscillations on the $\nu_e$ beam component.
In the results shown in Figs.~3 and 4 we have imposed $P_{ee} =1$, i.e. we have fixed
the $\nu_e$ flux at its non-oscillated value.  Now we relax such a condition allowing for values of $P_{ee} <1$.
In Fig.~5, the (red) dashed curve represents the upper bounds obtained in such more general situation. We see that, for large values of $|U_{e4}|^2$, there are appreciable deviations from the particular case $P_{ee} = 1$. Larger values of the appearance mixing angle $\theta_{\mu e}$ are now allowed by the fit. More precisely, from Fig.~5 we derive the upper bound $\sin^2 2\theta_{\mu e} \lesssim 1.7 \times 10^{-2}$,  which is a factor $\sim3/2$ bigger than that found in the case
 $P_{ee} = 1$ ($\sin^2 2\theta_{\mu e} \lesssim 1.2 \times 10^{-2}$) and an overall factor $\sim 3$ weaker than that derived using the 2-flavor approximation ($\sin^2 2\theta_{\mu e} \lesssim 5.2 \times 10^{-3}$). 
The reasons of this behavior can be traced to the fact that the fit has now more flexibility and to the circumstance
that the number of $\nu_e$ events measured by OPERA (and also by ICARUS) is appreciably lower than the theoretical (non-oscillated) background prediction. When including in the fit the possibility of having $P_{ee}<1$,  a large non-zero value of $|U_{e4}|^2$ is preferred, since this suppresses the background prediction and provides a better agreement with the observations. In this case, larger values of the $\nu_\mu \to \nu_e$ signal are naturally permitted by the fit and, as a consequence, bigger values of $\theta_{\mu e}$ are allowed.%

\section{Conclusion}

The two long-baseline experiments ICARUS and OPERA have recently performed sterile neutrino searches using the $\nu_\mu \to \nu_e$ measurements. Both collaborations have presented upper bounds on the effective appearance mixing angle $\theta_{\mu e}$ obtained with analyses which make use of an effective 2-flavor description. We have shown that a consistent treatment of the results must include genuine 4-flavor interference effects, which develop on the  long distances involved in the CNGS setup. Our quantitative study shows that their inclusion weakens the upper bounds on  $\theta_{\mu e}$ approximately by a factor of two. 
We have also pointed out that, in a 4-flavor scheme, the sterile-induced $\nu_e$ disappearance 
is of high relevance. Its inclusion in the data analysis leads to a further weakening of the upper bounds on 
$\theta_{\mu e}$, which overall are relaxed by a factor of three with respect to those obtained in the effective 2-flavor description.  In conclusion, the 4-flavor effects that we have investigated have a substantial impact on the data interpretation. Therefore, they should be  included in any accurate analysis.

\section*{Acknowledgments}
We  acknowledge support from the EU through the Marie Curie  
Fellowship ``On the Trails of New Neutrino Properties" (PIEF-GA-2011-299582) and
partial support from the FP7 ITN ``Invisibles'' (PITN-GA-2011-289442).

\bibliographystyle{h-physrev4}

\end{document}